\documentclass[]{article}
\usepackage{emulateapj}

\usepackage{graphicx}

\advance \voffset by 1.00cm\relax

\def\msun{{\rm ~M}_{\odot}}
\def\rsun{{\rm ~R}_{\odot}}

\def\mpy{{\rm ~M}_{\odot} {\rm ~yr}^{-1}}

\begin{document}

\title{IC10~X-1/NGC300~X-1: the very immediate progenitors of BH-BH binaries}

\author{Tomasz Bulik\altaffilmark{1,2},
        Krzysztof Belczynski\altaffilmark{1,3}, 
        Andrea Prestwich\altaffilmark{4}}

\affil{
     $^{1}$ Astronomical Observatory, University of Warsaw, Al. Ujazdowskie 4, 00-478 Warsaw,
            Poland\\
     $^{2}$ Nicolaus Copernicus Astronomical Center, Bartycka 18, 00-716 Warsaw, Poland\\
     $^{3}$ Center for Gravitational Wave Astronomy, University of Texas at
            Brownsville, Brownsville, TX 78520, USA\\
     $^{4}$ Harvard-Smithsonian Center for Astrophysics, Cambridge, MA 02138, USA\\
      tb@astrouw.edu.pl, kbelczyn@nmsu.edu, andreap@head.cfa.harvard.edu }

\begin{abstract}
We investigate the future evolution of two extragalactic X-ray binaries:
IC10~X-1 and NGC300~X-1. Each of them consists of a high mass BH ($\sim
20-30 \msun$) accreting from a massive WR star companion ($\gtrsim 20 \msun$), 
and both are located in low metallicity galaxies. We analyze the current state 
of the systems and demonstrate that both systems will very quickly ($\lesssim 
0.3$ Myr) form close BH-BH binaries with the short coalescence time ($\sim 3$ Gyr) 
and large chirp mass ($\sim 15 \msun$). 
The formation of BH-BH system seems unavoidable, as {\em (i)} WR companions
are well within their Roche lobes and they do not expand so no Roche lobe
overflow is expected, {\em (ii)} even intense WR wind mass loss does not
remove sufficient mass to prohibit the formation of the second BH, {\em (ii)} 
even if BH receives the large natal kick, the systems are very closely bound
and are almost impossible to disrupt.   
As there are two such immediate BH-BH progenitor systems within $2$ Mpc and as the 
current gravitational wave instruments LIGO/VIRGO (initial stage) can detect such 
massive BH-BH mergers out to $\sim 200$ Mpc, the empirically estimated detection 
rate of such inspirals is $R=3.36^{+8.29}_{-2.92}$ at the $99\%$ confidence level.
If there is no detection in the current LIGO/VIRGO data (unreleased year of $s6$ run), 
the existence of these two massive BH systems poses an interesting challenge. 
Either the gravitational radiation search is not sensitive to massive 
inspirals or there is some fundamental misunderstanding of stellar evolution 
physics leading directly to the formation of BH-BH binaries. 
\end{abstract}


\section{Introduction}

The interferometric gravitational wave observatories LIGO and VIRGO
have already reached their design sensitivities and both are undergoing
further improvements to reach the advanced sensitivity stage.
The most promising sources of gravitational waves that these experiments
are looking for are coalescences of compact objects. Among these
most attention has been paid to double neutron star systems (NS-NS).
There is an observational evidence of their existence and their merger rates
seem to warrant detection with the advanced interferometric experiments.
The double black hole binaries (BH-BH) and black hole neutron star 
binaries (BH-NS) have received less attention in the rate prediction calculations. 
There are several reasons for that: the direct detection
of such systems in electromagnetic domain is difficult. From the
theoretical point of view formation of such systems is not easy, as they
have to pass through an unstable mass transfer phase which is not easy to 
survive for typical black hole masses of around $10$M$_\odot$ (Belczynski et al. 2007).
However, it was shown recently that in low metallicity environment this
obstacle can be overcome and that formation rates of binary black holes can be 
quite large (Belczynski et al. 2010a).

The recent advances in X-ray instrumentation allow a study 
of X-ray binaries in the Local Group galaxies. 
IC10~X-1 has already been discovered in the ROSAT data 
by Brandt et al (1997). Bauer \& Brandt (2004) have found 
an X-ray variability of IC10 X-1 in a short Chandra observation.
Clark \& Crowther (2004) analyzed the possible optical counterparts 
of  IC10~X-1(Crowther et al. 2003), and argued that it is a $35$M$_\odot$ WNE star.
Subsequent longer Chandra observation lead to the discovery 
of X-ray periodicity (Prestwich et al. 2006). 
Prestwich et al. (2007) analyzed the X-ray and optical data 
 of IC10~X-1, and found that it contains a black hole of a mass at least $23$M$_\odot$
in a binary with a $\approx 35$M$_\odot$ companion.
This result has been recently confirmed by 
Silverman \& Filipenko (2008), who measured precisely the 
amplitude of the radial velocity of the companion.
NGC300 X-1 a system similar to  IC10 X-1 is the binary NGC300 X-1 (Crowther
et al. 2007). Crowther et al. (2010) measured precisely the radial velocity 
amplitude in NGC300~X-1 and showed that it contains a $20$M$_\odot$ black hole
accreting from a $26$M$_\odot$ Wolf-Rayet (WR, or naked helium) star. The 
orbital periods in both systems are similar.

Apparently, black holes of stellar origin can reach much larger mass than previously
thought. Although such high mass BHs are already fully explained by current 
evolutionary models (Belczynski et al. 2010b). Additionally, these most
massive stellar BHs can be found in binaries with very massive companions. 
In this paper we analyze the future binary evolution of IC10~X-1 and NGC300~X-1 
using the {\tt StarTrack} binary evolution code and we demonstrate that,
regardless of evolutionary uncertainties, these systems will form close high 
mass BH-BH binaries. Such BH-BH binaries formed at low metallicity were
already predicted, on theoretical grounds, to be the first detectable sources for 
gravitational radiation instruments like LIGO and VIRGO (Belczynski et al. 2010a).

\section{Model}

\subsection{Evolutionary code}

Our population synthesis code, {\tt StarTrack},  was initially developed to 
study double compact object mergers in the context of GRB progenitors
(Belczynski, Bulik \& Rudak 2002b) and gravitational-wave inspiral sources 
(Belczynski, Kalogera, \& Bulik 2002a). In recent years {\tt 
StarTrack} has undergone major updates and revisions in the physical treatment 
of various binary evolution phases, and especially the mass transfer phases. 
The new version has already been tested and calibrated against observations and 
detailed binary mass transfer calculations (Belczynski et al.\ 2008a), and has 
been used in various applications (e.g., Belczynski \& Taam 2004; Belczynski et 
al.\ 2004; Belczynski, Bulik \& Ruiter 2005; Belczynski et al. 2006; Belczynski 
et al.\ 2007). The physics updates that are most important for compact object 
formation and evolution include: a full numerical approach for the orbital evolution 
due to tidal interactions, calibrated using high mass X-ray binaries and open 
cluster observations, a detailed treatment of mass transfer episodes fully 
calibrated against detailed calculations with a stellar evolution code, updated 
treatment of mass transfer and common envelope phases, and the latest 
determination of the natal kick velocity distribution for neutron stars 
(Hobbs et al.\ 2005). The kicks for black holes are decreased proportionally
to the amount of fall back during core-collapse/supernova explosion. For
most massive stars ($M_{\rm zams} \gtrsim 40 \msun$ forming massive black holes 
without a supernova explosion (see Fryer \& Kalogera 2001) we assume no
natal kick. 

The most recent update, employed in this study, concerns wind mass loss from
massive stars. In particular interest here are mass loss rates from massive 
naked helium stars. For WR stars we adopt
\begin{equation}
(dM/dt) = 10^{-13} L^{1.5} \left({Z \over Z_\odot}\right)^{m} \mpy
\label{wind13}
\end{equation}
which is a combination of the Hamann \& Koesterke (1998) wind rate estimate
that takes into account WR wind clumping (reduced winds), and Vink \& de Koter
(2005) wind $Z$-dependence who estimated $m=0.86$ for WR stars.
Using the above estimate, along with other updated mass loss rates, we were
able to recover masses of most massive black holes in different galaxies 
(Belczynski et al. 2010b). 
The other helium star properties (e.g., radii, luminosities and lifetimes)
are adopted from Hurley, Pols \& Tout (2000) who employed detailed evolutionary
calculations for WR stars presented later by Pols \& Dewi (2002).
For the full description of the population synthesis code we refer the reader 
to Belczynski et al. (2008a).

\subsection{Host galaxy metallicity}

IC10 is a barred irregular galaxy in the Local Group at the distance
between 600 and 800 kpc (Saha et al. 1996; Sakai, Madore \& Freeman, 1999). 
It is undergoing a very rapid star formation and has a very high number of 
Wolf Rayet stars. IC10 has low metallicity: Lequeux et al. (1997) estimates 
it to be $Z=0.15 Z_\odot$, but later studies by Massey et al. (2007) place it 
somewhere between the values for LMC and SMC. Thus we conservatively adopt 
the value of $Z=0.3 Z_\odot$. IC10 is a galaxy with a large star formation 
rate and it contains more than a hundred WR stars (Massey \& Holmes, 2002)

The metallicity of  the NGC300 galaxy has been measured by 
Urbaneja et al. (2005). The galaxy exhibits some metallicity gradients
and at the location of NGC300~X-1 it is $\log(O/H)+12\approx 8.44$
which corresponds to the $Z=0.6 Z_\odot$ (Crowther et al. 2010).

\section{Results}

\subsection{The Future Evolution of IC10~X-1}

At present IC10~X-1 has an orbital period $34.93$ hr, black hole mass
is estimated to be $23-33 \msun$, while its companion is a helium star of a
mass $17-35 \msun$. The system is an eclipsing X-ray source with X-ray 
luminosity of $2 \times 10^{38}$ erg s$^{-1}$ an the lower limit on
inclination was placed at $78$ deg (Silverman \& Filipenko 2008).

The fate of the WR star is mainly set by the wind mass loss. We consider 
three cases for the IC10~X-1 describing the current state of the systems: 
case {\em (a)} where the BH mass is $23 \msun$ and the the WR star mass is $17 \msun$; 
case {\em (b)} where the BH mass is $28 \msun$ while the WR star mass is $25 \msun$; 
and case {\em (c)} where the BH mass is $33 \msun$ while the WR star mass is $35 \msun$. 
The evolution of the WR stars with the metallicity $Z=0.3 Z_\odot$ (IC10) is
followed (see the top panel of Figure~1). 
The wind mass loss rate depends strongly on the initial mass of the star and 
on its metallicity. In each case we assume that the measured mass corresponds 
to the initial, unevolved state of the WR star. In the case {\em (a)} 
the $17 \msun$ WR star looses $3 \msun$, in the case {\em (b)} the $25 \msun$ 
it looses $5 \msun$, while in the case {\em (c)} the $35 \msun$ star looses 
$6\msun$ over its entire lifetime. Due to the mass loss from
the WR component the orbit expands slightly (by $\sim 2 \rsun$) and the
period increases to $\sim 40$ hr. Throughout its evolution the massive WR
star, for any value of the adopted mass, is always well within its Roche 
lobe. For the intermediate mass of $25 \msun$, the radius of WR component is 
$R_{\rm WR} \sim 1-2 \rsun$ while the Roche lobe radius is $R_{\rm roche} 
\gtrsim 7 \rsun$. The Roche lobe volume filling factor is of the order of  
$f_{\rm fill} \equiv (R_{\rm WR}/R_{\rm roche})^3 \sim 0.01$. The WR
component is filling only $1\%$ of its Roche lobe and there is no chance of
Roche lobe overflow in this system, therefore the only mass transfer
proceeds via stellar wind as calculated in our model. 

The WR star eventually undergoes the core-collapse. The WR star, in
each case, is so massive that the initial star mass must have been above 
$M_{\rm zams} = 40 \msun$ (helium star mass is on average about $1/3$ of the 
initial star mass, e.g., Hurley et al 2000). According to  Fryer \& Kalogera 
(2001) calculations of core collapse such massive stars form BHs through
direct collapse, in other words entire star ends up in the BH as the
explosions are extremely weak for such high masses. However, we allow for
$10\%$ mass loss in neutrinos in our calculations. 
In each case a BH is formed, see Figure~1, and its mass varies from 
$14 \msun$ for the case {\em (a)} through $18 \msun$ for the case {\em (b)} to 
$26 \msun $ in the case {\em (c)}. 
Since there is no (or almost no) mass loss in the core-collapse, the BH most
likely receives no (or only small) natal kick.

In all cases the binary survives the formation of a second BH, and thus 
the IC10~X-1 evolution leads to the formation of a BH-BH binary. 
The mass loss in neutrinos induced a small eccentricity $e \approx 0.04$,
however the orbit remains mostly unchanged. The chirp mass of the newly 
formed binary varies from $15 \msun$ for the model {\em (a)} through 
$20 \msun$ for model {\em (b)} and up to $26 \msun$ in the case {\em (c)}.
In each case the binary mass and the size of the orbit make it merge in
relatively short time: $2.6$ Gyr {\em (a)}, $1.8$ Gyr {\em (b)} and $1.2$ Gyr 
{\em (c)}.

\subsection{The Future Evolution of NGC3000~X-1}

NGC300~X-1 has a period of $32.3$h, very close to that of IC10~X-1.
Crowther et al. (2010) report that the mass of the WR companion is likely 
to be $26 \msun$ which implies the mass of the BH to be $20 \msun$. However,
if other stars contribute to the measured optical flux than the WR star 
mass can be $15 \msun$ and the implied  BH mass is $14.5 \msun$.
We will refer to the latter estimate as case {\em (a)} while the former 
one with more massive BH will be denoted as case {\em (b)}. The evolution of 
the WR star at the metallicity appropriate for its location within NGC300  
($Z=0.6Z_\odot$) is calculated. The results are shown in the bottom panel of 
Figure~1. In the case {\em (a)} the $15 \msun$ WR star looses 
$3 \msun$, and forms an $11 \msun$ BH. In the case {\em (b)} the WR star 
looses $8 \msun$ in the wind, and forms a $16 \msun$ BH. Note that mass loss 
in the case of NGC300~X-1 is relatively higher than that of IC10~X-1, as the
former system host galaxy has more metal rich stars and therefore wind mass 
loss is more efficient. System orbit expands slightly (by $\sim 2-4 \rsun$) 
and the period increases to $\sim 37-47$ hr for case {\em (a)} and {\em
(b)}, respectively. Throughout its evolution the massive WR star 
($R_{\rm WR} \sim 1-2 \rsun$), for any value of the adopted mass, is always 
well within its Roche lobe ($R_{\rm roche} \gtrsim 6 \rsun$). 
The Roche lobe volume filling factor is very small ($f_{\rm fill} \sim 0.01$) 
and there is no chance for Roche lobe overflow. 

At the time of core collapse the WR component is massive enough to form a 
BH through direct collapse. The mass of the second BH is $11 \msun$ {\em
(a)} or $16 \msun$ {\em (b)}. The system acquires small eccentricity 
($e \approx 0.04$). The merger time of the binary BH system is $4.0$Gyr 
in the case {\em (a)} and $3.8$Gyr in the case {\em (b)}. The chirp mass of 
the BH-BH binary is between $11 \msun$ {\em (a)} and $15\msun$ {\em (b)}. 
Thus NGC300~X-1 is another example of a system that will evolve to form a 
merging BH-BH.

\subsection{Estimate of the coalescence rate}

The Chandra or XMM sensitivity to detect X-ray binaries like IC10~X-1 or 
NGC300~X-1 extends to distances beyond the Local Group. However the 
sensitivity to fully analyze such binaries is limited by the possibility
of obtaining spectroscopic orbits. This in turn is limited by the 
brightness and spectral properties of the WR stars. The absolute brightness 
of a massive WR star is about $M_v\approx -5$, and a spectroscopic orbit 
can be measured for stars with the apparent magnitude down to $m_v \approx 21$. 
Thus a detailed spectroscopic orbit of a WR star can be obtained up to a 
distance of $\approx 2$~Mpc. Let us assume that such binaries could be 
detected up to the distance of $r_{s}$, and only a fraction $\Omega_s$ of 
the sky has been searched for such systems therefore we can estimate the 
volume in which they are detectable as $V_s=\Omega_s r_{s}^3/3$.
 
The entire  sky has not been surveyed for such binaries, yet we can assume
that almost all the sky has been searched for such binaries so $\Omega_s=4\pi$. 
This is a conservative assumption, i.e. it underestimate the formation rate 
of binary BHs, since we overestimate the volume surveyed for such binaries 
so far. The lifetime of the IC10~X-1 in the X-ray bright phase (accretion
from intense wind of WR companion) is not longer than $t_{IC10}\approx 0.3 $Myr, 
while for the NGC300~X-1 it is smaller than $t_{NGC300} \approx0.2$ Myr (see
Fig.~1). The current formation rate density of merging compact object 
binaries from IC10~X-1 and NGC300X-1 like systems can be estimated as 
\begin{eqnarray}
\rho_{IC10} \approx V_{s}^{-1}t_{IC10}^{-1}\\
\rho_{NGC300} \approx V_{s}^{-1}t_{NGC300}^{-1}\\
\rho= \rho_{IC10}+\rho_{NGC300}
\end{eqnarray}
Assuming that the star formation rate is constant and noting that the merger 
times of the BH-BH binaries described above are significantly smaller than the 
Hubble time this is also the estimate of the current BH-BH merger rate density. 
A detailed calculation, presented in the Appendix, leads to the estimate of the 
merger rate density: $\rho= 0.36^{+0.50}_{-0.26}$Mpc$^{-3}$Myr$^{-1}$ at the 
$90\%$ confidence level.

In order to estimate the detection rate in current gravitational wave detectors 
we must take into account the different chirp masses of the two binaries. In   
order to be conservative we will only consider the low mass cases: {\em (c)} for 
IC10~X-1 with $M^{IC10}_{chirp}=15\msun$ and {\em (b)} for NGC300~X-1 with 
$M^{NGC300}_{chirp}=11\msun$.

We assume that the current LIGO sensitivity allows it to detect a NS-NS binary 
with a chirp mass of $1.2\msun$ to a distance of $r_{NSNS}=18$\,Mpc. This is the 
sky averaged horizon. The sensitivity range depends on the chirp mass $M_{chirp}$ 
of a given binary and scales as $M_{chirp}^{5/6}$. The detection rate of BH-BH
inspirals is a sum of the rates originating in IC10~X-1 and in NGC300~X-1 like 
binaries:
\begin{eqnarray}
{\cal R}_{IC10}={4\pi\over 3}{r_{NSNS}}^3 \rho_{IC10}\left({M^{IC10}_{chirp}\over 1.2 \msun}\right )^{5/2}\\
{\cal R}_{NGC300}={4\pi\over 3}{r_{NSNS}}^3 \rho_{NGC300}\left({M^{NGC300}_{chirp}\over 1.2 \msun}\right )^{5/2}\\
{\cal R}= {\cal R}_{IC10}+{\cal R}_{NGC300}
\label{rate}
\end{eqnarray}
We present the probability density distribution of the total rate {\cal R}
as well as the contributions from each type of binary in Figure~2.
We have calculated the confidence intervals 
for 
the total rate, and they are
$R=3.36^{+2.44}_{-1.62}$ at the 68\% confidence level,
$R=3.36^{+4.55}_{-2.32}$ at the 90\% confidence level, and
$R=3.36^{+8.29}_{-2.92}$ at the 99\% confidence level.

\section{Discussion}

\subsection{Potential Caveats}

There are two crucial points that lead to determination of the
mass of the BH in both of the systems: (i) the estimate of the mass
of the companion WR star, and (ii) the estimate of its orbital velocity.
In the case of (i), the WR star mass, we have used the lowest 
estimate consistent with observations in both cases.
 Higher masses of the
WR star imply higher BH masses and  only increase the estimate of the
coalescence
rate. Thus the uncertainty from the WR mass estimate is not crucial.
The second point (ii), the estimate of the orbital velocity may however
be overestimated by a systematic effect.  It has been suggested by
van Kerkwijk (1993) that the wind could be highly ionized except for
the region shadowed by the star. This ionized wind model was
originally developed
in the context of Cyg X-3.
In this model large velocities of the lines in the spectrum
in this model  originate from the wind rather then from the orbital velocity
of the WR star. Such an effect would lead to overestimate of the
orbital velocity
and overestimate of the mass of the mass of the BH.
Such an ionized wind model predicts blueshifted spectra at the moment of
the X-ray eclipse, and can be distinguished by simultaneous X-ray and optical
observations. Such observations are not yet available for the
binaries considered here.
 Moreover, such a model would lead to a different shape
of the radial velocity curve, so detailed time resolved  spectroscopy
of the system might be helpful. One needs to note that such an ionized
model may be difficult to apply to IC10-X-1 and NGC300 X-1 because of  their much
longer orbital periods than the orbital period of Cyg X-3 which is 4.8h.

Our conclusion on the lack of ongoing or future Roche lobe overflow is based 
on the fact that stellar models indicate that massive WR stars are very compact 
$R_{\rm WR} \sim 1-2 \rsun$ (e.g., Pols \& Dewi 2002) while both systems under 
consideration are relatively wide, with Roche lobe radii of WR components of 
the order of $R_{\rm roche} \gtrsim 6-7 \rsun$. 
However, it was claimed that some specific physical conditions (iron opacity 
peak) within WR stars may lead to the radial inflation of the outer atmosphere 
(e.g., Ishii, Ueno \& Kato 1999). This effect was recently examined in detail 
by Petrovic, Pols \& Langer (2006) and it was found that for sub-solar
metallicities and realistic calculations (with wind mass loss) the WR stars
with mass $\sim 10-30 \msun$ remain compact ($<2 \rsun$). 
Radii of massive WR stars are very difficult to be observationally determined 
even if bolometric luminosity and temperature is known (e.g., Hamman et al. 1995;
Moffat \& Marchenko 1996). Due to strong optically thick winds WR stars appear 
larger than they most likely are. For example, the radius of WR component in
NGC300~X-1 was observationally determined $\sim 5-7 \rsun$ (Crowther et al.
2010), while stellar models indicate much smaller value. 

We have assumed that the currently observed WR mass is its initial mass. If in fact, 
the WR star was more massive initially and if by now it has gone through part of its
evolution (as it most likely did) it means that the mass loss from now on will be 
lower (less time left till core-collapse) than we have estimated. Therefore, if 
anything, we underestimate the final mass of WR component and the BH mass it will 
form. Our mass loss rates are on the high side and thus we may be additionally 
underestimating the mass of the second BH. For example, in case of IC10~X-1 the 
mass loss is estimated at the level of $\sim 10^{-5} \mpy$ (Crowther \& Clark 2004) 
while we employ the wind mass loss rates in the range $1-6 \times 10^{-5} \mpy$ 
depending on the adopted mass of WR component (see Fig.~1). This is even more clear 
in the case of NGC300~X-1 for which mass loss rate is determined to be 
$\sim 5 \times 10^{-6} \mpy$ (Crowther et al. 2010) while we employ much higher 
rates $\sim 5 \times 10^{-5} \mpy$.

Based on high WR star mass in both systems, we have followed Fryer \& Kalogera 
(2001) to infer that second BHs in IC10~X-1 and NGC300~X-1 will form through
direct collapse of entire star to a BH. Since there is no mass loss in such
a case we have assumed zero natal kick in both cases. This holds true if the natal
kicks are connected with asymmetry in mass ejection during supernova explosion. 
This has some observational support in the fact that most massive BHs in our 
Galaxy, that are believed to have formed through direct collapse, show no sign of 
natal kick (Mirabel \& Rodrigues 2003; Dhawan et al. 2007; Martin et al. 2010). 
If, for some reason, these massive BHs received a kick as large as 
observed for Galactic NSs ($\sim 200-300$ km s$^{-1}$; e.g., Hobbs et al.
2005) both systems would survive and still form close (but rather eccentric)
BH-BH binaries with coalescence time below Hubble time. 
The relative orbital velocity in the pre-explosion binary for both IC10~X-1 and 
NGC300X-1 is $\approx 600$km\,s$^{-1}$. Since even large natal kick is not likely 
to exceed the orbital velocity the binaries are expected to survive (e.g.,
Kalogera 1996). 

The estimate of the initial LIGO detection rate is very high and it is actually 
quite conservative. Including the more accurate fraction of the sky that has been 
surveyed, leads to a  decrease of $\Omega_s$ and increase of the expected rate.  
The estimated range $r_s$ to obtain spectroscopic orbits of such binaries
has been calculated neglecting the potential effects of extinction. Decreasing this 
range increases the expected coalescence rate density. The expected 
rate depends on the inverse of assumed lifetime in the X-ray phase, and we have 
chosen the maximum possible values for the WR lifetime in equation (4).

Both IC10~X-1 and NGC300~X-1 are very young systems with current age shorter than 
$t_{\rm evol} \approx 10$ Myr (it is an evolutionary lifetime of a $M_{\rm ZAMS}=20 
\msun$ star). Therefore, both systems were formed very recently in Universe
with rather low star formation. The merger time for both systems was estimated at 
$\sim 2-4$ Gyr. If the production of binaries like IC10~X-1 and NGC300~X-1 is in
anyway proportional to the star formation rate (as it seems logical), it must have 
been higher $2-4$ Gyr ago, as the star formation increases with redshift all the 
way to redshifts of $z \approx 2$ ($\approx 10$ Gyr ago; e.g., Strolger et al. 2004).  
Since we have assumed that the star formation rate (SFR) is constant as a function
of redshift at the level that corresponds to the current low SFR (the very young age 
of both binaries) we have again underestimated the formation efficiency of
similar binaries and the predicted detection rates are lower limits with respect to 
the SFR. 
Moreover, the median metallicity of galaxies was lower a few billion years ago 
when the currently merging systems have formed (e.g., Pei, Fall \& Hauser 1999; 
Young \& Fryer 2007). This could also bring the current coalescence rate up
as it appears that the formation of binaries with massive BHs occurs in low 
metallicity host galaxies. 

On the other hand it is obvious that our arguments are based on just two objects 
and therefore are subject to small number statistics and one should treat our
results only as an indication of possibility of existence of a large population 
of merging massive BH-BH binaries. 
In summary, the value of the rate presented here should be considered as a lower 
limit, while keeping in mind that our arguments and calculations are based
only on two objects.

\subsection{Conclusions}

We have demonstrated that the future evolution of the binaries IC10~X-1 and NGC300~X-1
will lead to the formation of BH-BH binaries with a merger time of $2-4$ Gyr. Such 
systems are representative of a population that should be detectable by the 
interferometric detectors like LIGO and VIRGO. Our estimate of the formation rate
density of such systems is $ 0.36$Mpc$^{-3}$Myr$^{-1}$, which implies the detection 
rate of their coalescences by current LIGO and VIRGO of $3.36$\,yr$^{-1}$. This means 
that such a merger should be found in the already gathered LIGO data! The absence of a 
massive BH-BH inspiral signal in the $s6$ LIGO and VIRGO data, at their current 
sensitivities, will imply astrophysically challenging limits on the coalescence rates 
of massive BH-BH binaries.

This estimate is similar to the one in the recently published theoretical study of
the BH-BH formation in low metallicity environment (Belczynski et al. 2010a). In that 
paper the authors find that low metallicity environment greatly enhances formation 
rate of massive BH-BH binaries. This is due to the fact that low metallicity binaries 
are much more likely to survive the common envelope stage and form close BH-BH systems. 
Survival of the common envelope stage was shown to be a key issue in the BH-BH 
formation (Belczynski et al. 2007). Belczynski et al. (2010a) have provided population 
synthesis prediction of the BH-BH detection rate under assumption that there is a 
significant population of low metallicity galaxies ($50\%$; Panter et al. 2008) in the
local Universe. They have provided a rate estimate for two models, one in
which they do not allow for the suppression of BH-BH formation due to common envelope
mergers (model A) and one in which such a suppression is physically motivated and 
allowed (model B). The BH-BH detection rates for the current LIGO were
calculated at such a high level for model A ($\sim 5$ yr$^{-1}$) that Belczynski et al. 
(2010a) excluded this model as unrealistic in anticipation on no detection 
in the just accomplished $s6$ LIGO run\footnotetext{For model B the rates
obtained by Belczynski et al. (2010a) are much lower ($\sim 0.05$ yr$^{-1}$)
and are consistent with no detection in $s6$ LIGO run.}. However, it seems that the 
existence of IC10~X-1 and NGC300~X-1 seems to support such high detection rate. 

In conclusion, if in fact there is no gravitational radiation signal from
inspiraling massive BH-BH binaries in the last $s6$ LIGO run, either the
current search is insensitive to high chirp mass binaries or there is some
fundamental misunderstanding of the last stages of evolution leading directly 
to the formation of BH-BH mergers from the binaries like IC10~X-1 and
NGC300~X-1.

\acknowledgements
We would like to thank Chris Fryer, Kris Stanek, Aleks Schwarzenberg-Czerny and 
Richard O'Shaughnessy for their helpful comments. Authors acknowledge support from 
MSHE grants N N203 302835 (TB, KB), N N203 404939 (KB) and NASA Grant
NNX09AV06A to the Center for Gravitational Wave Astronomy, UTB (KB).

\appendix

Let us assume that the formation rate per unit time per unit volume of massive X-ray 
binaries similar to IC10 X-1 is $\rho$. Then the expected number of such binaries in 
a volume V,  given that they are X-ray active for a time T, is $\lambda=\rho V T$. 
The probability of observing one object is then given by the Poisson distribution:
\begin{equation}
P(1|\lambda) = \lambda e^{-\lambda}
\end{equation}
We are interested in measurement of $\rho$, thus we can use the Bayes theorem to 
obtain the probability of the rate given a single observation
\begin{equation}
P(\lambda|1)= P(1|\lambda)P(\lambda)/P(1)
\end{equation}
where $P(\lambda) $ is the prior probability, and $P(1)$ can be treated as the 
normalization of the resulting probability distribution.   We assume a flat prior  
$P(\rho)=\rm const$. Given the observed volume  $V={4\pi\over 3} R_s^3$, where 
$R_s\approx 2 Mpc$, and the time $T_{IC10}=0.2Myrs$, we obtain the probability 
distribution of $\rho_{IC10}$, the formation rate of IC10 like systems:
\begin{equation}
{dP\over d\rho_{IC10}} = A^2 \rho e^{-A\rho}
\end{equation}
where $A=V T_{IC10}$. Analogously for NGC300 X-1 like systems we have the 
probability distribution of the formation rate:
\begin{equation}
{dP\over d\rho_{NGC300}} = B^2 \rho e^{-B\rho}
\end{equation}
where $B= V T_{NGC300}$, and $T_{NGC300}=0.3$Myrs.  We can now calculate the 
probability density of the total formation rate of binaries like IC10~X-1 and 
NGC300~X-1, $\rho=\rho_{IC10}+\rho{NGC300}$:
\begin{equation}
{dP\over d\rho} =\int \int  d\rho_{IC10}d\rho_{NGC300} 
\delta(\rho-(\rho_{IC10}+\rho{NGC300})){dP\over d\rho_{IC10}}{dP\over d\rho_{NGC300}}.
\end{equation}
A simple calculation yields:
\begin{equation}
{dP \over d\rho} ={A^2B^2\over (A-B)^3} 
\left[ e^{-\rho B} (\rho(A-B)-2) - e^{-\rho A} (\rho(B-A)-2)\right]
\end{equation}.  
We identify the formation rate with the coalescence rate, which can be 
justified since the coalescence time of each binary is smaller than the Hubble 
time. 

In order to find the probability distribution of the LIGO  detection rate coming 
from each of the binary we notice that the probability density of the detection 
rate is connected with the merger  rate via ${\cal R}_{IC10}  = V^{GW}_{IC10} \rho_{IC10}$, 
where 
\begin{equation}
V^{GW}_{IC10}= {4\pi\over 3} r_{NSNS}^3 \left(
{ {\cal M}_{IC10} \over {1.2 M_\odot}  }
\right)^{5/2}
\end{equation}
is the volume in which BH-BH from IC10~X-1 like binaries are detectable,
$r_{NSNS}=18$Mpc 
is the current sky averaged sensitivity distance for detection by LIGO of binaries 
with the chirp mass of $1.2M_\odot$ and $ {\cal M}_{IC10}$ is the chirp mass of the 
BH-BH binary that will form from  IC10~X-1. For the chirp mass we conservatively 
assume the minimum value  $ {\cal M}_{IC10}=15 M_\odot$. The probability density of 
the detection rate ${\cal R}_{IC10}$ is
\begin{equation}
{dP\over d{\cal R}_{IC10}} = {dP\over d\rho_{IC10}} {d\rho_{IC10}\over d{\cal R}_{IC10}}.
\end{equation}
 Analogously the LIGO probability density of detection rate of BH-BH binaries originating 
in NGC300~X-1 like binaries is
\begin{equation}
{dP\over d{\cal R}_{NGC300}} = {dP\over d\rho_{NGC300}} {d\rho_{NGC300}\over d{\cal R}_{NGC300}},
\end{equation}
and we use $ {\cal M}_{NGC300}=11 M_\odot$ for the calculation. A simple calculation 
shows that the probability distribution densities are
\begin{equation}
{dP\over d{\cal R}_{IC10}} = a^2 \rho e^{-b\rho}
\end{equation}
\begin{equation}
{dP\over d{\cal R}\rho_{NGC300}} = b^2 \rho e^{-b\rho}
\end{equation}
where  $a=0.749$yr, and b=$1.07$yr. The calculation of the probability density of the 
total rate ${\cal R} =  {\cal R}_{IC10}+{\cal R}\rho_{NGC300} $ follows along the same 
lines as above:
\begin{equation}
{dP\over d{\cal R}}
 =\int \int  d{\cal R}_{IC10}d\rho_{NGC300} 
\delta({\cal R}-(
{\cal R}_{IC10}+
{\cal R}_{NGC300}
))
{dP\over d{\cal R}_{IC10}}
{dP\over d{\cal R}_{NGC300}}.
\end{equation}
leads to the result:
\begin{equation}
{dP \over d{\cal R}} ={a^2b^2\over (a-b)^3} 
\left[ e^{-{\cal R} b} ({cal R}(a-b)-2) - e^{-{\cal R} a} ({cal R}(b-a)-2)\right]
\end{equation}.

\clearpage

\begin{figure}
\includegraphics[width=0.9\columnwidth]{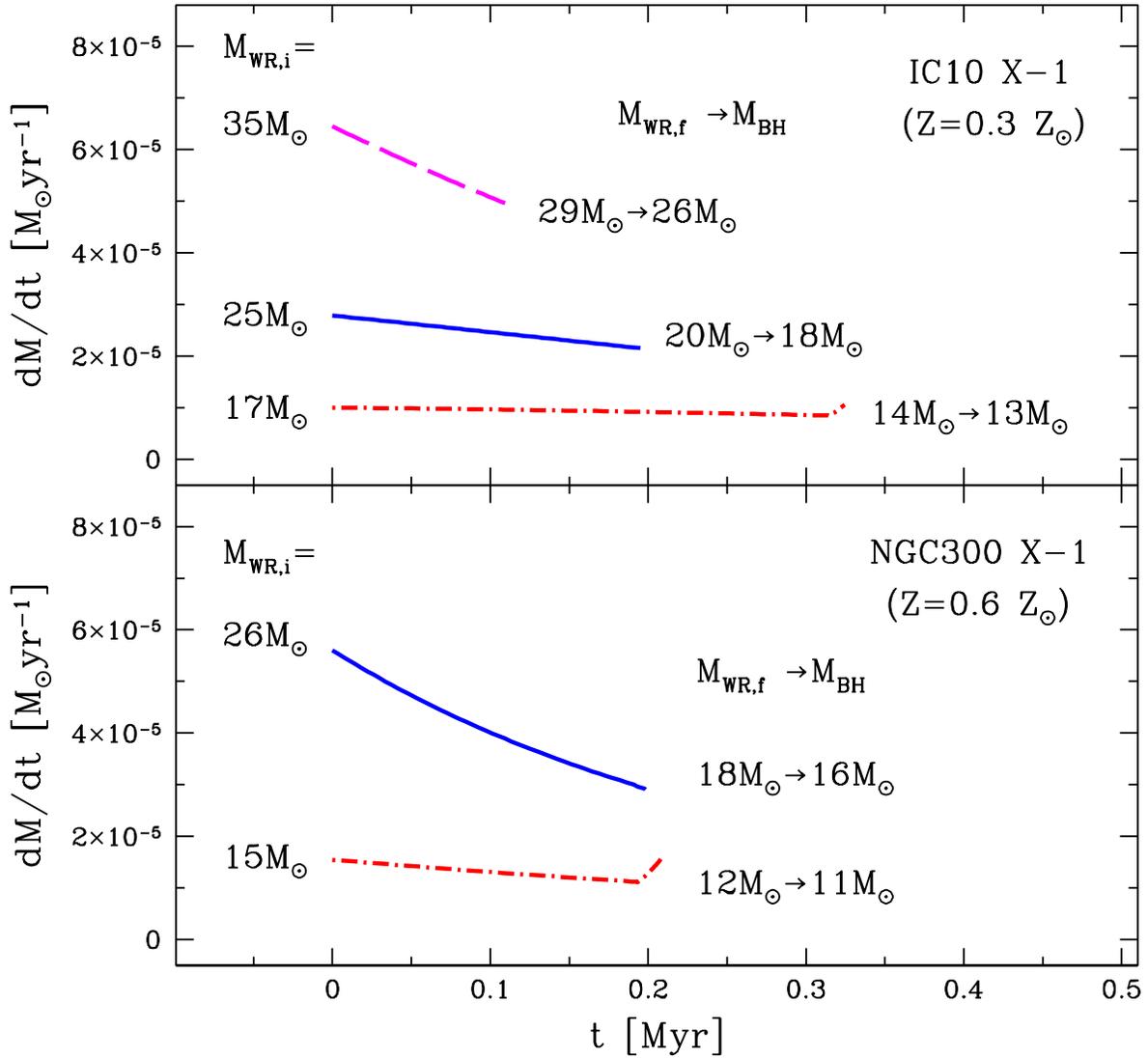}
\caption{The mass loss rates from  WR stars as a function of age 
in the two systems considered in the paper. 
The top panel corresponds to IC10~X-1 and the bottom panel shows the case of
NGC300~X-1.
Each line is labeled by the initial mass of the star on the left hand side.
On the right hand side we show the final mass of the star, and the mass of the 
compact object (BH) formed as a result of the evolution.}
\end{figure}

\begin{figure}
\includegraphics[width=0.6\columnwidth]{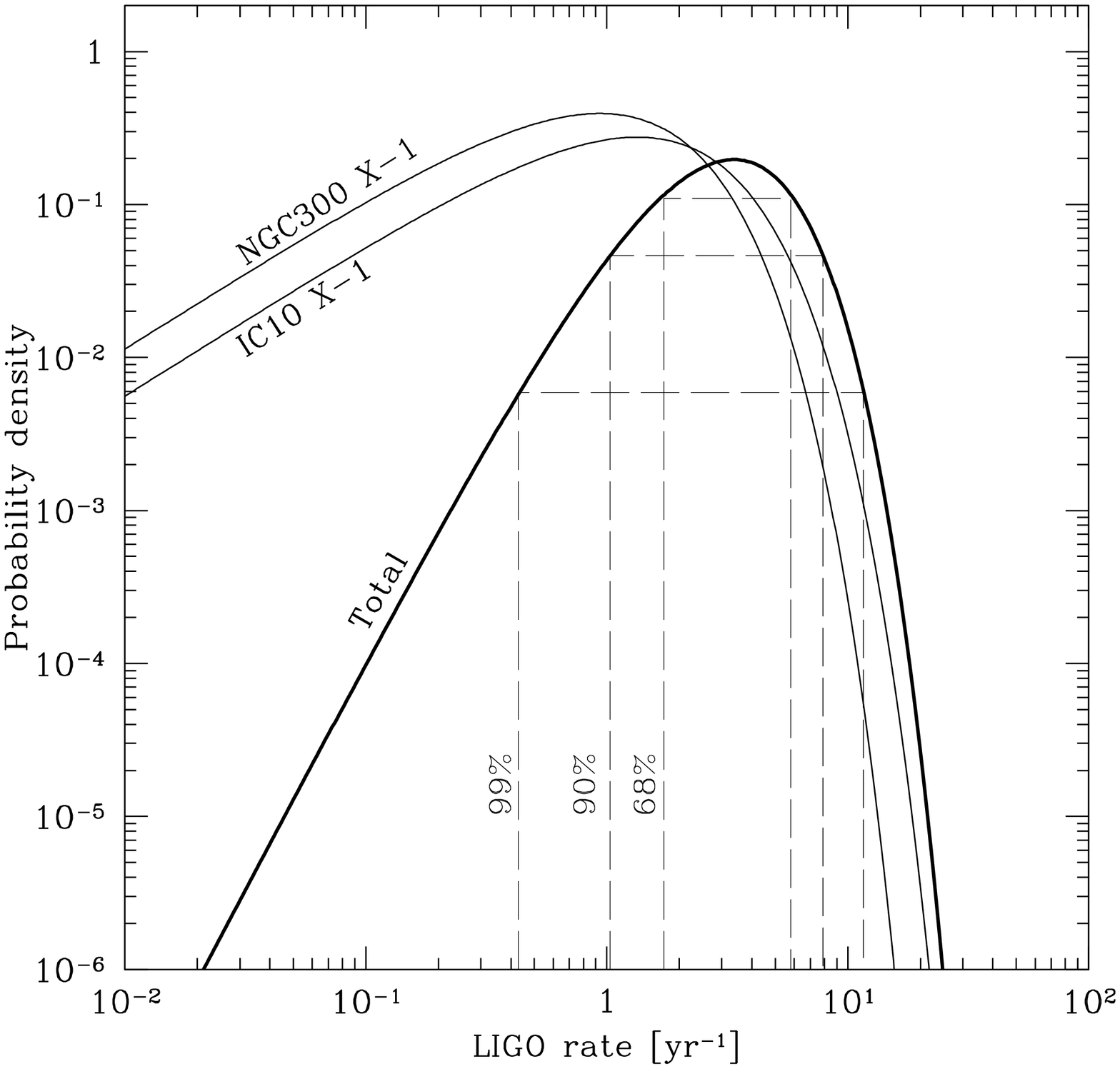}
\caption{The probability distributions of the total detection rate in LIGO - thick 
solid line, along with the contributions from each of the binaries: IC10~X-1
and NGC300~X-1 - thin solid lines. The dashed lines denote the 
levels and intervals corresponding to the ranges containing 68\%, 90\%, and 99\%
of probability. The empirical estimate of the detection rate of BH-BH binaries for 
the initial LIGO is $R=3.36^{+8.29}_{-2.92}$ at the $99\%$ confidence level.
}
\end{figure}


\begin{references}
\reference{} Abadie et al.\ 2010, CQG, 27, 173001
\reference{} Bauer, F., \& Brandt, W.\ 2004, \apj, 601, L67 
\reference{} Belczynski, K., Kalogera, V., \& Bulik, T.\ 2002a, \apj, 572,
             407
\reference{} Belczynski, K., Bulik, T., \& Rudak, B.\ 2002b, \apj, 571, 394
\reference{} Belczynski, K., Kalogera, V., Zezas, A., \& Fabbiano, G.\ 2004,
             \apj, 601, L147
\reference{} Belczynski, K., \& Taam, R.\ 2004, \apj, 616, 1159
\reference{} Belczynski, K., Taam, R., Kalogera, V., Rasio, F., \& Bulik, T.\ 
             2007, \apj, 662, 504 
\reference{} Belczynski, K., Bulik, T., \& Ruiter, A.\ 2005, \apj, 629, 915
\reference{} Belczynski, K., Perna, R., Bulik, T., Kalogera, V., Ivanova,
             N., \& Lamb, D.Q.\ 2006, \apj, 648, 1110
\reference{} Belczynski, K., Taam, R., Kalogera, V., Rasio, F., \& Bulik,
             T.\ 2007, \apj, 662, 504
\reference{} Belczynski, K., Kalogera, V., Rasio, F., Taam, R., Zezas, A., 
             Bulik, T., Maccarone, T., \& Ivanova, N. \ 2008a, \apjs, 174, 223
\reference{} Belczynski, K., Taam, R., Rantsiou, E., \& van der Sluys, M.\
             2008b, \apj, 682, 474
\reference{} Belczynski, K., Dominik, M., Bulik, T., O'Shaughnessy, R., 
             Fryer, C., \& Holz, D.\ 2010a, ApJ 715, L138
\reference{} Belczynski, K., Bulik, T., Fryer, C., Ruiter, A., Valsecchi, F., 
             Vink, J., \& Hurley, J.\ 2010b, ApJ, 714, 1217
\reference{} Brandt, W., Ward, M., Fabian, A., \& Hodge, P.\ 1997, 
             \mnras, 291, 709 
\reference{} Bulik, T., Belczynski, K., \& Rudak, B.\ 2004, \aap, 415, 407 
\reference{} Clark, J., \& Crowther, P.\ 2004, \aap, 414, L45
\reference{} Crowther, P., Carpano, S., Hadfield, L., \& Pollock, A.\ 
             2007, \aap, 469, L31 
\reference{} Crowther, P., Drissen, L., Abbott, J., Royer, P., \& Smartt, S.\ 
             2003, \aap, 404, 483 
\reference{} Crowther, P. A.; Barnard, R.; Carpano, S.; Clark, J. S.; Dhillon, V. S.;
Pollock, A. M. T., 2010, MNRAS, 403, 41 
\reference{} Dewi, J., \& Pols, O.\ 2003, \mnras, 344, 629
\reference{} Dhawan, V., et al.\ 2007, ApJ, 668, 430
\reference{} Fryer, C., \& Kalogera, V.\ 2001, \apj, 554, 548 
\reference{} Hamann, W.-R., Koesterke, L., \& Wessolowski, U.\ 1995, \aap, 299, 151 
\reference{} Hamann, W.-R., \& Koesterke, L.\ 1998, \aap, 335, 1003 
\reference{} Hobbs, G., Lorimer, D., Lyne, A., \& Kramer,
             M.\ 2005, \mnras, 360, 974
\reference{} Humphreys, R, \& Davidson, K.\ 1994, \pasp, 106, 1025 
\reference{} Hurley, J., Pols, O., \& Tout, C.\ 2000, \mnras, 315, 543 
\reference{} Ishii, M., Ueno, M., \& Kato., M.\ 1999, PASJ, 51, 417
\reference{} Ivanova, N., Belczynski, K., Kalogera, V., Rasio, F., \&
             Taam, R. E.\ 2003, \apj, 592, 
\reference{} Kalogera, V.\ 1996, \apj, 471, 352 

\reference{} Kudritzki, R., \& Reimers, D.\ 1978, \aap, 70, 227
\reference{} Lequeux, J., Peimbert, M., Rayo, J., Serrano, A., \& Torres-Peimbert, S.\ 
             1979, \aap, 80, 155 
\reference{} Lorimer, D.\ 2005, NATO ASIB Proc.~210: The Electromagnetic 
             Spectrum of Neutron Stars, 161 
\reference{} Martin, R., Tout, C., \& Pringle, J.\ 2010, MNRAS, 410, 1514
\reference{} Massey, P., \& Holmes, S.\ 2002, \apj, 580, L35 
\reference{} Massey, P., McNeill, R., Olsen, K., Hodge, P., Blaha, C., Jacoby, 
             G., Smith, R., \& Strong, S.\ 2007, \aj, 134, 2474
\reference{} Mirabel, F., \& Rodrigues, I.\ 2003, Science, 300, 1119
\reference{} Moffat, A., \& Marchenko, S.\ 1996, \aap, 305, L29
\reference{} Nieuwenhuijzen, H., \& de Jager, C.\ 1990, \aap, 231, 134 
\reference{} Nugis, T., \& Lamers, H.\ 2000, \aap, 360, 227 
\reference{} Orosz, J., et al.\ 2007, Nature, 449, 872
\reference{} Panter B., et al.\ 2008, MNRAS, 391, 1117
\reference{} Pei, Y., Fall, M., \& Hauser, M.\ 1999, \apj, 522, 604
\reference{} Petrovic, J., Pols, O., \& Langer, N.\ 2006, \aap, 450, 219 
\reference{} Pols, O., \& Dewi, J.\ 2002, \pasp, 19, 233 
\reference{} Prestwich, A., Kilgard, R., Carpano, S., Saar, S., Page, K., Roberts, A., 
             Ward, M., \& Zezas, A.\ 2006, The Astronomer's Telegram, 955, 1       
\reference{} Prestwich, A., et al.\ 2007, \apj, 669, L21
\reference{} Sadowski, A., Belczynski, K., Bulik, T., Ivanova, N., Rasio, F., \& 
             O'Shaughnessy, R.\ 2008, ApJ, 676, 1162
\reference{} Saha, A., Hoessel, J., Krist, J., \& Danielson, G.\ 1996, \aj, 111, 197 
\reference{} Sakai, S., Madore, B., \& Freedman, W.\ 1999, \apj, 511, 671 
\reference{} Silverman, J., Filipenko, A., 2008, ApJ 678, L17
\reference{} Strolger, L., et al.\ 2004, \apj, 613, 200
\reference{} Timmes, F., Woosley,  S., \& Weaver, T.\ 1996, \apj, 457, 834 
\reference{} Urbaneja, M., et  al.\ 2005, \apj, 622, 862 
\reference{} Webbink, R. F.\ 1984, \apj, 277, 355
\reference{} van Kerkwijk, M.~H.\ 1993, \aap, 276, L9 
\reference{} Vassiliadis, E., \& Wood, P.\ 1993, \apj, 413, 641 
\reference{} Vink, J., \& de Koter, A.\ 2005, \aap, 442, 587 
\reference{} Young, P. \& Fryer, C.L.\ 2007, \apj, 670, 584



\end{references}
\end{document}